\documentstyle[12pt]{article}
\newcommand{\beq}{\begin{equation}}
\newcommand{\eeq}{\end{equation}}
\newcommand{\beqn}{\begin{eqnarray}}
\newcommand{\eeqn}{\end{eqnarray}}
\newcommand{\bea}{\begin{eqnarray}}
\newcommand{\eea}{\end{eqnarray}}

\newcommand{\lqcd}{\Lambda_{QCD}}

\newcommand{\Dslash}{\slash \hspace{-8pt} D}

\newcommand{\Qslash}{\slash \hspace{-6.5pt} Q}

\newcommand{\nn}{\nonumber}
\begin{document}
\pagestyle{empty} 
\vspace{-0.6in}
\begin{flushright}
ROME1-1204/98 \\
TUM-HEP-309/98 \\
\end{flushright}
\vskip 1cm
\centerline{\large{\bf{MODEL INDEPENDENT DETERMINATION OF}}}
\centerline{\large{\bf{THE SHAPE FUNCTION FOR INCLUSIVE $B$ DECAYS AND}}}
\centerline{\large{\bf{OF THE STRUCTURE FUNCTIONS IN DIS}}}
\vskip 1.4cm
\centerline{U.~Aglietti$^1$, M.~Ciuchini$^{2}$, G.~Corb\`o$^1$,}
\centerline{E.~Franco$^1$, G.~Martinelli$^{1}$, L.~Silvestrini$^3$}
\vskip  0.4cm
\centerline{\small $^1$ Dipartimento di Fisica, Universit\`a ``La Sapienza" 
and INFN,}
\centerline{\small Sezione di Roma, P.le A. Moro, I-00185 Rome, Italy.}
\centerline{\small $^2$ INFN, Sezione Sanit\`a, V.le Regina Elena 299,
 Rome, Italy.}
\centerline{\small $^3$ Physik Department, Technische Universit\"at M\"unchen,}
\centerline{\small D-85748 Garching, Germany.}
\vskip 1.5cm
\abstract{
We present  a   method to compute, by numerical  simulations of 
lattice QCD, the inclusive semileptonic  differential decay rates  of heavy 
hadrons and the structure functions which occur in deep inelastic 
scattering.  The method is based  on first principles
and does not require any model assumption. It allows 
the prediction of the 
differential rate in $B$ semileptonic decays for values of the recoiling
hadronic mass $W \sim \sqrt{M_{B}\lqcd}$, which is in the relevant region
to extract $\vert V_{ub} \vert$ from the end-point of the lepton spectrum
in inclusive decays.}
\eject
\pagestyle{empty}\clearpage
\setcounter{page}{1}
\pagestyle{plain}
\newpage 
\pagestyle{plain} \setcounter{page}{1}
\section{Introduction}
\label{sec:intro}
In this paper  we propose a method  to compute, on the lattice, the shape
function $f(k_+)$~\cite{fkr,fkn} which enters  the calculation of 
the inclusive differential semileptonic decay
rate of heavy hadrons.  The knowledge of $f(k_+)$ is a fundamental
ingredient for the
extraction of $\vert V_{ub} \vert$ from the end-point of the lepton spectrum.
The same function also enters the calculation of the photon spectrum in
radiative $B$ decays. Furthermore,
the same method can be applied to the calculation
of the structure functions of Deep Inelastic Scattering (DIS). 
Our approach does not require any model assumption and is based on  standard
techniques which are used to compute the hadronic matrix elements in
 lattice  QCD. In particular, we show that 
the use of the Euclidean space-time, which is unavoidable in lattice
calculations, is not an obstacle to the calculation of $f(k_+)$
in the deep inelastic region.
\par Our main result  is that, from the study of suitable combinations of 
lattice Green functions, we can obtain the quantity
\begin{equation} \label{eq:mainre}
G(t,\vec Q)= \frac{e^{- \sqrt{\vec Q^2} t}}{2 \sqrt{\vec Q^2}}
\int^{M_B}_0 \ \tilde f(k_+) \ e^{-  k_+ t }  \ , 
\end{equation}
where $M_B$ is the mass of the decaying hadron (the $B$ meson in the
example considered in this paper), $t$ is a time distance and $\vec Q$ a spatial
 momentum that we can
inject in the Green functions computed in the numerical simulations.
By varying $t$ and  $\vec Q$, we can unfold the integral and extract $\tilde f(k_+)$.
\par The function $\tilde f(k_+)$  in eq.~(\ref{eq:mainre}) 
differs from the usual shape function $f(k_+)$  introduced in
refs.~\cite{fkr,fkn}, as will be explained in the following.
In terms of $\tilde f(k_+)$, the differential semileptonic decay rate
can be written as
\bea \label{eq:dslr0}
\frac{d \Gamma}{dE_{\ell}} &\equiv& \int^{M_B}_{0} dk_{+} \ \tilde f(k_{+}) \
\frac{d \Gamma_{PM}}{dE_{\ell}}(m_b^*,E_\ell)\nn \\ &=&\vert V_{u b} 
\vert^{2} \frac{G_{F}^{2} }{12 \pi^{3} }\ E_{\ell}^{2} 
 \int^{M_B}_{0} dk_{+} \ \tilde f(k_{+}) \
\Theta( m^{*}_{b} - 2 E_{\ell} ) \left[ 3 m^{* \, 2}_{b} - 4 m^{*}_{b} E_{\ell} 
\right] \, , \eea
 where $m^{*}_{b}= M_B -  k_{+}$. 
The advantage of using $\tilde f(k_+)$ is that 
any reference to unphysical quantities, such as 
$\bar \Lambda$ or the quark mass $m_b$, disappears  in the expression above and 
 the differential rate is written in terms of hadronic quantities only. 
\par The  paper is organised as follows. In section~\ref{sec:inclusive}
we review the main formulae for the total and differential semileptonic decay 
rates as they can be derived using the Operator Product Expansion 
(OPE);
 $f(k_+)$ is introduced in this section.
In  sec.~\ref{sec:formulae} we present the basic formalism needed 
to explain our idea.  We discuss,  as a prototype,  the case of   the
shape function which enters  the differential decay rate, and then extend 
the method  to the computation of  the structure functions.
 We express these quantities  in 
terms of suitable $T$-products of local operators
and discuss their  continuation to the Euclidean space-time.
In  sec.~\ref{sec:implementation}, we describe the implementation of the 
method in lattice calculations and discuss the systematic errors
which may arise in actual numerical simulations, due
to the present limitations in computer resources.
\section{Inclusive decay rates}
\label{sec:inclusive}
This  section contains a summary of the main results obtained in the 
literature for  semileptonic and radiative 
inclusive decays. We present below the main formulae for the total and 
differential rates which can be derived using the OPE~\cite{fkr}--\cite{hqet}. 
This will allow us to
introduce the definition of the shape function and of its moments.

The idea that inclusive decay rates of hadrons containing heavy quarks can be computed 
in the parton model is quite old and was used, for example,
 in ref.~\cite{cabma} to predict the charmed-hadron lifetimes. To account
for bound state effects, the partonic  calculation was 
subsequently improved by the introduction of a
  phenomenological model, called  
the ``spectator model" ~\cite{pieta,accmm}, which has
been and continues to be
 extensively used to extract $\vert V_{cb}\vert $ and $\vert V_{ub}\vert$ from 
 inclusive semileptonic decays. A noble theoretical framework
for the spectator model was then provided  by the the Wilson OPE 
applied to the inclusive decays of heavy hadrons~\cite{fkr}--\cite{hqet}.
 The  expansion parameter is not necessarily
the inverse heavy-quark mass $1/m_Q$, rather  the inverse of the
energy release $1/W$
of the process at hand.  In a large region of the available phase space
 $W$  is  of the order of $m_Q$.
In this case,  under the hypothesis of quark-hadron duality,
the operator-product expansion is expected to give 
accurate predictions for the decay widths, expressed in terms of  few 
non-perturbative parameters. In particular, at 
lowest order in $1/m_{Q}$, the expression of the decay widths 
derived with the OPE  coincides with  that 
obtained in the parton model calculation.

Let us recall how this works for the inclusive semileptonic
processes $\bar B \to X_{c,u} \ell \bar \nu_{\ell}$.
Using the OPE, the total semileptonic decay rate is given by~\cite{bigi,mw} 
\bea \label{eq:slr}  \Gamma(\bar B \to X_{c,u}  \ell \bar \nu_\ell ) =
\vert V_{(c,u)b} \vert^2 \frac{G_F^2 m_b^5}{192 \pi^3} 
\left[  \left(1 + \frac{ \lambda_{1}}{2 m^{2}_{b}} \right)
 C_{0}(x_{c,u})  -  \frac{9 \lambda_{2}}{2 m_b^2} C_{1}(x_{c,u})
+ \dots \right] \, , \eea
where we have neglected  QCD radiative corrections and  phase space 
effects due to the mass of the final charged lepton.
The dots stand for higher-order perturbative
 and/or power corrections. We
 have omitted  to write explicitly the
terms of  ${\cal O}(1/m_b^3)$   because they are expected
to give  negligible contributions to $\Gamma(\bar B \to X_{c,u} \ell \bar \nu_\ell )$.
 $ C_{0,1}(x_{c,u})$  are known phase-space factors which depend
on the ratios of the final-quark  masses to $m_b$, 
$x_{c,u}=m^{2}_{c,u}/m^{2}_b$~\cite{bigi,mw}.
In eq.~(\ref{eq:slr}), the  term  which coincides with 
 the parton model  result
\bea \label{eq:slrpm}  \Gamma_{PM}(\bar B \to X_{c,u}  \ell \bar \nu_\ell ) =
\vert V_{(c,u)b} \vert^2 \ \frac{G_F^2 m_b^5}{192 \pi^3} \
 C_{0}(x_{c,u}) \eea
corresponds to the insertion of the leading, dimension-three operator
$\bar b b$ appearing in the OPE.  
In the heavy quark effective theory (HQET) at fixed velocity $v$, 
this is a conserved  
operator ($ \bar b b  \to \bar h_{v} h_{v} + \dots$)  satisfying
the following normalisation condition
\beq  \label{eq:sd}\langle \bar B(v) \vert \bar h_{v} 
h_{v } \vert \bar B(v) \rangle = 1 \, .\eeq
The hadron state $\vert \bar B(v) \rangle$ is normalised to $v^0$ instead of
the usual relativistic normalisation of $2 E$ because this is more convenient
for the heavy quark expansion.
The corrections of ${\cal O}(1/m_{b})$ vanish since the only possible 
dimension-four operator, $\bar b i\, \Dslash  b$, can be reduced to
$\bar b b$  by using the equations of motion.  
The kinetic energy of the heavy quark in the $B$ meson, 
$\lambda_{1}$, and the chromomagnetic moment of the heavy quark, 
$\lambda_{2}$, correspond to  matrix elements of local,  dimension-five
operators 
appearing in the OPE~\cite{hqet}. Written in terms of the fields of 
the HQET, these operators are given by
\bea \label{eq:las} \lambda_{1} &=& \langle 
\bar B(v) \vert \bar h_{v} (i D)^{2}
h_{v } \vert \bar B(v) \rangle \nn \\
 \lambda_{2} &=& \langle \bar B(v) \vert \bar h_{v} \sigma_{\mu\nu}
 i D^{\mu}  i D^{\nu} h_{v } \vert \bar B(v) \rangle \, .\eea
Thus, up to ${\cal O}(1/m_{b}^{3})$ the problem is reduced to the 
calculation of the matrix elements of the two local operators appearing
in eq.~(\ref{eq:las}). For these matrix elements several theoretical 
estimates exist~\footnote{ We are not concerned here with the precise 
definition of $\lambda_{1,2}$  when QCD corrections are taken into 
account.}. A recent review of these estimates  can be found, for example,
 in ref.~\cite{jeru}.
\par Unfortunately, in order to determine $\vert V_{ub}\vert $ from the
lepton spectrum, one has to study the differential distribution 
and perform cuts  to suppress decays involving charmed particles in 
the final state. For the differential distribution, the relevant scale of the
 OPE  is the squared momentum of the hadronic 
system recoiling against the lepton pair, $Q_b= m_{b} v - q$, where $q$ 
is the momentum of the lepton system. 
 By approaching the end-point of the 
electron spectrum  $W=\sqrt{Q_b^{2}}$ becomes small. 
\par In the region
where  $v \cdot Q_b$ is of ${\cal O}(M_{B})$ and $W \sim \sqrt{M_{B} 
\Lambda_{QCD}}$  we can still use the parton model by introducing a 
``shape function'', analogous to the distribution function of the
spectator model~\cite{accmm}, corresponding to a modified OPE. This is analogous 
to deep inelastic scattering, where one can introduce non-perturbative 
distribution functions for the light-cone component of the quark 
momenta.
In terms of the shape function $f(k_{+})$ the differential distribution 
is given by
\bea \label{eq:dslr}
\frac{d \Gamma}{dE_{\ell}} &\equiv& \int^{\bar \Lambda}_{-m_b} dk_{+} \ f(k_{+}) \
\frac{d \Gamma_{PM}}{dE_{\ell}}(m_b^*,E_\ell)\nn \\ &=&\vert V_{u b} 
\vert^{2} \frac{G_{F}^{2} }{12 \pi^{3} }\ E_{\ell}^{2} 
 \int^{\bar \Lambda}_{-m_b} dk_{+} \ f(k_{+}) \
\Theta( m^{*}_{b} - 2 E_{\ell} ) \left[ 3 m^{* \, 2}_{b} - 4 m^{*}_{b} E_{\ell} 
\right] \, , \eea
 where $m^{*}_{b}= m_{b} + k_{+}$ and
\beq \label{eq:fkp}  f(k_{+}) = \langle \bar B(v) \vert \ \bar h_{v}
\delta(k_{+} - i D_{+}) h_{v} \ \vert \bar B(v) \rangle \, .\eeq
$D_{+} = n \cdot D$, with $n^\mu = Q_b^\mu /(v \cdot Q_b)$. 
Note that in the region where $W \sim \sqrt{M_{B} 
\Lambda_{QCD}}$, $n^{2} \sim 
\Lambda_{QCD}/M_{B} \sim 0$.
\par Finally, if we consider the extreme region where $W \sim \Lambda_{QCD}$, 
the partonic picture  
breaks down. In this region the rate is dominated by few single 
states or resonances and the appropriate description can only be 
obtained by predicting the rate as a  sum over the lowest exclusive 
channels.
\par  At the lowest order in $1/m_b$, the integration of   $d \Gamma / d 
E_{\ell}$ 
over  $E_{\ell}$ in eq.~(\ref{eq:dslr}) gives the total rate. It also gives
some of the ${\cal O}(1/m_b^2)$ corrections appearing in eq.~(\ref{eq:slr}).
For the following discussion, it is convenient to introduce the moments
of $f(k_{+})$, defined as 
\beq \label{eq:mom} {\cal M}_{n} = \int_{-m_b}^{\bar \Lambda} dk_{+} k^{n}_{+} f(k_{+}) \, . \eeq
The first few moments are known:
the leading contribution,  given by ${\cal M}_{0} =1$,
is the 
normalisation of $f(k_{+})$ and is fixed by the normalisation of the 
conserved scalar density operator of eq.~(\ref{eq:sd}); the second
moment ${\cal M}_{1}$, related to the correction of ${\cal O}(1/m_b)$, 
vanishes  because  there are no operators contributing at this order; 
the correction of ${\cal O}(1/m_b^2)$
is given by  ${\cal M}_{2}=-\lambda_{1}/3$. The contribution of the 
heavy-quark chromomagnetic moment is not included in the shape 
function.   \par
In the case of the integrated rate, 
the contributions coming from higher moments of the shape functions
are  related to operators of dimension larger than five.
For this reason, they are suppressed by higher powers in 
$1/m_{b}$ and can be safely ignored.  For the differential
distribution of eq.~(\ref{eq:dslr}) near the end-point, instead, 
all  moments ${\cal M}_{n}$  become of the same order in $1/m_{b}$  and cannot be 
neglected~\footnote{ A more extensive discussion of this
point will be presented in section~\ref{sec:formulae}.}.  
The knowledge of the full shape function $f(k_{+})$, and 
not only of the first few moments, is 
then needed in this region.  This is  analogous to  what happens
for the structure 
functions in DIS  processes in the region 
where the Bjorken variable  $x \to 1$. \par 
 So far, several models for the shape function have been 
proposed~\cite{fkr,fkn,accmm,models}. In particular, 
one can show that the old ACCMM  model is equivalent, up to a process-dependent
redefinition of the heavy quark mass~\cite{fkr}, to 
a specific choice of the shape function.
It remains true, however, that the extraction of $\vert V_{ub}
\vert$  from the  experimental measurements of  the end-point of the lepton spectrum 
using  these models is ``model dependent'' and consequently affected
by a systematic error which is difficult to estimate.
 
\section{The shape function $\tilde f(k_+)$}
\label{sec:formulae}
 In this section, we show that it is possible
to determine  the full shape function, and not only a few moments of 
it (as for example  $\lambda_{1}$~\cite{gms}),  by numerical simulations on the lattice.
The shape function can be obtained from the  time and momentum dependence  of 
suitable Euclidean Green functions of the same type  as those which 
are currently used on the lattice to compute  hadronic matrix elements 
of local operators. We show that, in the  inelastic region where 
the knowledge of the full shape function is necessary, this 
can be done  in spite of the fact that lattice simulations are 
performed in the Euclidean space-time.  This is new since, as   
discussed below and in  
ref.~\cite{panic96}, the use of the Euclidean space-time, 
instead of the Minkowsky one, seems to prevent the possibility of computing 
inclusive quantities on the lattice~\footnote{For exclusive processes
involving more than one particle in the final state, the problem is usually 
referred to as the Maiani-Testa no-go theorem \cite{mt}.}.
We also show that, by using the same approach, it is possible to compute the 
structure functions of DIS, and not only few moments of it, as  
done so far~\cite{sm1}--\cite{schi}. 
\par A further advantage of our 
proposal is the following. Higher  moments of the shape function 
(and those of 
the structure functions) are related to  matrix elements of local 
operators of higher dimensions which are, in general, 
afflicted by power divergences when a hard cutoff is used (or 
renormalon ambiguities in dimensional regularisations). This makes the 
definition of the renormalised operators, which have finite matrix 
elements when the cutoff is removed, very 
problematic~\cite{smrenormalon}. With our method, instead, 
no renormalisation is needed as  we get directly the shape function from
the lattice correlation functions,
 as explained in sec.~\ref{sec:implementation}.
\par
We now give all the details of the derivation of our method.
\vskip 0.3 cm
The spin-averaged, differential semileptonic decay rate is given by
\beq \label{eq:diffe} \frac{d \Gamma}{ dq^{2} dE_{\ell} dE_{\nu_{\ell}} } =
\vert V_{(c,u)b} \vert^{2} \frac{G_{F}^{2} }{2 \pi^{3} }  \left[
W_{1} \ q^{2}    + 
 W_{2} \left( 2 E_{\ell} E_{\nu_{\ell}} -\frac{1}{2} q^{2} 
\right) + W_{3} \ q^{2} \left( E_{\ell} - E_{\nu_{\ell}} \right)  \right] \, .\eeq
with the Lorentz-invariant functions $W_{i}$ 
 defined by the  expansion of the hadronic tensor
\beq \label{eq:ht}
W^{\mu\nu} = (2\pi)^{3} \sum_{X} \delta^{4} ( p_{B} - q  - 
p_{X} )  \langle \bar B(v) \vert J^{\mu \ \dagger} \vert X \rangle
\langle X \vert J^{\nu } \vert \bar B(v ) \rangle \, ,\eeq
which can be written as 
\bea W^{\mu\nu} &=&  -g^{\mu\nu} W_1 + v^\mu v^\nu W_2  - i 
\epsilon^{\mu \nu \alpha \beta} q_\alpha v_\beta  W_3 \nonumber \\
&+& q^\mu q^\nu W_4 + (q^\mu v^\nu + q^\nu v^\mu) W_5 \, .\eea
In eq.~(\ref{eq:ht}), the weak current is defined as $J^{\mu}\equiv
\bar q(x) \gamma^{\mu}_{L} b(x) $, 
where $q(x)$ represents the field of the light final quark
and $\gamma^{\mu}_{L}= \gamma^{\mu}
(1-\gamma_{5})/2$.
Using the optical theorem,  the form factors $W_i$ are given by
\beq W_i = - \frac{1}{\pi} \mbox{Im}T_i \ , \eeq
where the $T_i$ are defined by the forward matrix element 
of the $T$-product of the two weak currents
\bea \label{eq:tpro21} T^{\mu\nu} &=& - i \int d^4 x \ e^{-i q \cdot x} \
\langle \bar B(v) \vert  T \left( J^{\mu \ \dagger}(x)
 J^{\nu }(0) \right) \vert \bar B(v ) \rangle  \nonumber \\
  &=&  -g^{\mu\nu} T_1 + v^\mu v^\nu T_2  - i 
\epsilon^{\mu \nu \alpha \beta} q_\alpha v_\beta   T_3 \nonumber \\
&+& q^\mu q^\nu T_4 + (q^\mu v^\nu + q^\nu v^\mu) T_5 \, .
\eea
Thus the problem is reduced to the calculation of the forward matrix element
of the $T$-product defined above. \par One could  think  
 that, with sufficient   computer resources
to work with  very small  values lattice spacing $a$ such that  
\beq \label{eq:condi}\Lambda_{QCD}\ll m_b  \ll \frac{1}{a}  \ , \eeq
it  would be  possible to compute directly $W^{\mu\nu}$ on the lattice 
using the $T$-product of  eq.~(\ref{eq:tpro21}).
This  is in general illusory.
The problem arises from the fact that numerical simulations
are necessarily performed in the  Euclidean space-time~\cite{panic96}. 
Let us illustrate where  the problem comes from. \par In the Minkowsky case,
for $t >0$, one has 
\bea &\,& W^{\mu \nu} = \frac{1}{\pi} \mbox{Im} \left[ i \ \int \ d^4x  \  e^{-i q \cdot x}
\  \langle \bar B(v) \vert J^{\mu \ \dagger} (x)\  J^\nu(0)
\vert \bar B(v)  \rangle
\right] \  \nn
\\ &=& \frac{1}{\pi} \mbox{Im} \Bigg[ i \ \sum_{X} \int^{+\infty}_0 dt \  
e^{i(p_{B}^0 - q^0 -E_X) t}
\ (2 \pi)^3 \ \delta^3( \vec p_{B} - \vec q-\vec p_X)\times\nn\\
&& \qquad\qquad\qquad
\langle \bar B(v) \vert J^{\mu\ \dagger }(0) \vert X \rangle \langle X \vert J^\nu(0) \vert
\bar B(v)  \rangle \Bigg] \nn \\  &=& 
\sum_X (2 \pi)^3 \delta^4(p_B - q -p_X)
 \langle \bar B(v) \vert J^{\mu \dagger} (0)\vert X \rangle \langle X \vert 
 J^\nu(0) \vert
\bar B(v)  \rangle  \ . \label{eq:tpro2} \eea
For $t \le 0$, the cut  corresponds to an intermediate state with
two $b$-quarks and a $\bar c$, which is not the process we are interested
in.
The $\delta$ function related to energy conservation 
is given by the integral over the time of an oscillating exponential.
In the  Euclidean case, instead, eq.~(\ref{eq:tpro2}) becomes
\beq W^{\mu \nu} \sim \ \sum_{X} \int \ dt \  
e^{(p_{B}^0 - i q^0 -E_X)t}
\ (2 \pi)^3 \ \delta^3( \vec p_B - \vec q-\vec p_X)
\langle  \bar B(v) \vert J^{\mu \ \dagger}(0)
\vert X \rangle \langle X \vert J^\nu(0) \vert
\bar B(v) \rangle  \ . \label{tproe} \eeq
Thus, at large time distances, the integral over $t$ is dominated by
the states with the smallest energy $E_X$, and not by those
which satisfy the energy conservation condition $E_X=p_B^0 -q^0$. This is
similar to  what happens in the case of exclusive
decays which was  discussed in ref.~\cite{mt}.
The same problem is also encountered  in other analytic approaches which study inclusive
cross-sections or decay rates in the Euclidean \cite{blok}. 
\par We now show that, in the deep inelastic limit,
the analytic continuation to the Euclidean space-time of the $T$-product 
of the currents in eq.~(\ref{eq:tpro21}) becomes very simple and that it is 
possible to extract the wave function by studying suitable Euclidean Green
functions.
\par  $T^{\mu \nu}$ can be written as 
\bea \label{eq:tmunue} T^{\mu\nu}&=& -i \int d^4 x \ e^{-i q \cdot x} \
\langle \bar  B(v) \vert  \bar b(x) \gamma^\mu_L S(x-0) \gamma^\nu_L b(0) 
 \vert \bar B(v ) \rangle  \nonumber \\ &=&
-i \int d^4 x \ e^{i(p_B- q )\cdot x} \
\langle \bar B(v) \vert  \bar b_v(x) \gamma^\mu_L S(x-0) \gamma^\nu_L b_v(0) 
 \vert \bar B(v ) \rangle  \nonumber \\ &=&  
-i \int d^4 x \  
\langle \bar B(v) \vert  \bar b_v(x) \gamma^\mu_L S_Q(x-0) \gamma^\nu_L b_v(0) 
 \vert \bar B(v ) \rangle \ , \eea 
where $S(x-0)$ is the final light-quark propagator;
$b(x) \equiv e^{-i p_B \cdot x} b_v(x)$ 
and $S_Q(x-0)= e^{i Q \cdot x}
S(x-0)$, with $Q=p_B -q= M_B v - q$.
Our definition of the heavy quark field $b_v(x)$, written   
in terms of the physical momentum of the meson $p_B$ instead of the quark
momentum $p_b$, is legitimate since the shape function
is only defined at the lowest  order of the heavy quark expansion.
The reason for this choice will be explained below. 
It is possible to expand  $S_Q(x-0)$ as follows (for clarity we neglect
 the mass of the light quark)
\bea \label{eq:expaSQ} S_Q(x-0) &=& \left(\frac {i}
 {\Qslash + i \Dslash+ i \epsilon}\right)_{x-0} =
i \left(\frac{\Qslash + i  \Dslash}
{ Q^2 + 2 i Q \cdot D - D^2 - 1/2 \sigma_{\mu\nu}
G^{\mu \nu}+i \epsilon }\right)_{x-0} \nonumber \\ &\sim& 
i \left(\frac{\Qslash}{ Q^2 + 2 i Q \cdot
D + i \epsilon }\right)_{x-0} \ . \eea
In   eq.~(\ref{eq:expaSQ}) we have kept only the leading terms
of the expansion in powers of $1/\sqrt{Q^2} = 1/W \sim 1/M_B$, and 
those which become leading near
the end-point of the lepton spectrum, i.e. where $W \sim \sqrt{M_B
\Lambda_{QCD}}$~\cite{fkr}--\cite{mw}.
Using  eq.~(\ref{eq:expaSQ}) one finds
\beq \label{eq:deff+}
T^{\mu \nu} = \frac{1}{2}
\ \left(Q^\mu v^\nu + Q^\nu v^\mu - g^{\mu\nu} Q \cdot v
+ i \epsilon^{\mu\nu \alpha \beta }Q_\alpha v_\beta \right)
\int_0^{M_B} dk_+ \ \frac{\tilde f(k_+)}{Q^2 - 2 v \cdot Q k_+ + i \epsilon}\ ,
 \eeq 
where the shape function $\tilde f(k_+)$ is defined through the relations
\beq \label{eq: momentD} (-1)^n
\langle \bar B(v) \vert \bar b_v \gamma^\nu (i D^{\mu_1}) \dots
(i D^{\mu_n}) b_v \vert \bar B(v) \rangle = {\cal M}_{n} v^\nu v^{\mu_1}
\dots v^{\mu_n}  + {\cal B}_n \delta^{\nu \mu_1} v^{\mu_2} \dots v^{\mu_n}
+ \dots \ . \eeq
The  moments ${\cal M}_n$ in the equation above  are given by
\beq {\cal M}_n = \int_0^{M_B} dk_+ \  k_+^n \tilde f(k_+) \label{eq:mmn} \eeq
 With our  choice of $Q$, written in terms of the $B$-meson momentum, 
and up to a trivial change in sign, the $k_+$ of  eq.~(\ref{eq:dslr}) is changed
into $ -k_+ +\bar \Lambda$, leading to eq.~(\ref{eq:dslr0}).  
Note that with our definition of $k_{+}$, 
$m_{b}^{*}=M_{B}-k_{+}$, which is in our opinion more physical:
the differential distribution is now expressed in terms of hadronic 
quantities only, without any reference to unphysical quantities such 
as the quark mass or $\bar \Lambda$.
\par 
The moments  ${\cal M}_n \sim \Lambda_{QCD}^n$ 
have been defined  in eq.~(\ref{eq:mmn}) and give rise to
terms  proportional to 
 ${\cal M}_n (v \cdot Q/Q^2)^n \sim (\Lambda_{QCD} M_B/W^2)^n$. 
Thus, their contribution to the rate is suppressed as $(\Lambda_{QCD}/M_B)^n$ 
when $W \sim M_B$,   whereas it becomes of
${\cal O}(1)$ in the region where $W \sim \sqrt{M_B \Lambda_{QCD}}$.
 The contributions proportional to ${\cal B}_n$ are subleading in
$1/M_B$ with respect to the corresponding ${\cal M}_n$
in all the physical region of interest, i.e. for all values of
$W$, including   $W \sim \sqrt{M_B \Lambda_{QCD}}$ (but not
in the elastic region where $W \sim \Lambda_{QCD}$). 
 Note that in order to derive 
eq.~(\ref{eq:deff+}), we have never used the HQET, but only the OPE, by
separating  the large frequency modes ($\sim W$) from the low energy ones
($\sim \Lambda_{QCD}$) and expanding in powers of $\Lambda_{QCD}/W$. 
For simplicity, without loss of generality, we consider in the
following the case of  a $B$ meson at rest, namely $v =(1,\vec 0)$. 
\par We now consider, for $t >0$, the  Fourier transform of $T^{\mu\nu}$
defined as
\beq \label{eq:tmnt} T^{\mu \nu}(t, \vec Q)= \int
 \frac{d Q_0 }{2 \pi}
e^{-i Q_0 t} T^{\mu \nu} \ .\eeq
In terms of the $T$-product of the currents, we have
\beqn T^{\mu \nu}(t, \vec Q) &\equiv &-i \int d^3 x \ e^{-i \vec Q \cdot \vec x} \
 \langle \bar B(\vec p_B=0) \vert J_v^{\mu \dagger} (\vec x, t) J_v^\nu(0) \vert
\bar B(\vec p_B=0)  \rangle  \nonumber
\\ \label{eq:pippo} &=& -i e^{-i M_B t} \int d^3 x \ e^{-i \vec Q \cdot \vec x} \
 \langle \bar B(\vec p_B=0) \vert J^{\mu \dagger} (\vec x, t) J^\nu(0) \vert
\bar B(\vec p_B=0)  \rangle ,\eeqn
where  $J_v^\mu(x) = \bar q(x) \gamma^\mu_L
 b_v (x) $. The factor $e^{-i M_B t}$ appearing in 
 eq.~(\ref{eq:pippo}) cancels the corresponding term in the
 three-dimensional Fourier transform of the correlator, so that
 the r.h.s. of this equation goes as $e^{- i Q_0 t}$, which is the expected
 behaviour for a hadronic system with energy $Q_0$. 

For $t \ge 0$, by closing the contour of the integration over $Q_{0}$
below the real axis,  it is straightforward to find
\bea \label{eq:tmnte} T^{\mu \nu}(t, \vec Q) &=&  - \frac{i}{2} 
\int_{0}^{M_B} dk_+ \ \tilde f(k_+) 
 \left( \bar Q^\mu \delta^{\nu 0}
+ \bar Q^\nu \delta^{\mu 0} - g^{\mu \nu} \bar  Q^0 - i \epsilon^{0 \mu \nu
\alpha } \bar Q_\alpha \right) \times \nonumber \\
&\, & \frac{e^{- i \left( k_+  +
\sqrt{ \vec Q^2} \right) t } }{2 \sqrt{ \vec Q^2}} \ ,\eea
where $\bar Q \equiv (Q_0^+, \vec Q)$, with 
\beq Q^+_0 = k_+ +\sqrt{k_+^2 + \vec Q^2} \sim k_+ + \sqrt{ \vec Q^2} \ .\eeq
 \par 
Using eq.~(\ref{eq:tmnt}) it is very easy to make the analytic continuation to the
Euclidean space-time
\bea \label{eq:nclb} & \, & W^{\mu\nu}(t, \vec Q) = - \frac{1}{\pi}
\mbox{Im} T_E^{\mu \nu}(t, \vec Q)  \\
&=&  \frac{1}{2\pi } \int_{0}^{M_B} dk_+ \ \tilde f(k_+) 
\frac{e^{- \left( k_+ +
\sqrt{ \vec Q^2} \right) t } }{2 \sqrt{ \vec Q^2}}
 \times  \left( \bar Q^\mu \delta^{\nu 0}
+ \bar Q^\nu \delta^{\mu 0} - g^{\mu \nu} \bar  Q^0 - i \epsilon^{0 \mu \nu
\alpha } \bar Q_\alpha \right)  \ . \nonumber \eea
By a suitable choice of the Lorentz components $\mu$ and $\nu$ of the currents
and of the spatial momentum $\vec Q$, one can isolate either 
an integral of the form  given in eq.~(\ref{eq:mainre}) or the following one
\beq \label{eq:mainre1}
G(t,\vec Q)= \frac{e^{- \sqrt{ \vec Q^2} t } }{2 \sqrt{ \vec Q^2}}
\int_{0}^{M_B} dk_+  \ \tilde f(k_+) \ e^{- k_+  t } 
 \  Q_0^+
\, . \eeq
By studying the time dependence of $G(t,\vec Q)$ at several values 
of $\vec Q$, we can unfold both the integral above and the one
in eq.~(\ref{eq:mainre}) and extract the shape 
function.
\par One may be surprised that the analytic continuation to the Euclidean
space-time is so simple, in spite of the argument made at the beginning of
the section. Indeed, also in the Minkowsky case, in order to apply the OPE, one has first
to expand  the $T$-product of the currents for large space-like momenta,
which is equivalent to work in the Euclidean, and then continue the resulting
expression to the kinematic region of interest. If this is possible,
which implicitly corresponds to  the assumption of  quark-hadron duality,
then the expansion has the very simple form of eqs.~(\ref{eq:deff+})
and (\ref{eq:tmnte}). In these equations,  the
singularities are those of a free particle propagator, the Wick rotation
of which is  straightforward and leads to the result given in 
eq.~(\ref{eq:nclb}). The contamination of   
 intermediate states with  small invariant masses, which would dominate
at large time distances, is suppressed by injecting
in the correlation functions
a large spatial momentum $\vec Q$, with $\sqrt{\vec Q^2} \gg \Lambda_{QCD}$.
\par We now show that the same method can be used to compute the
structure functions of deep inelastic scattering. The starting formula is, 
as before, the $T$-product of two currents.
We have
\bea &\,& T^{\mu \nu} = -i \  \int \ d^4x  \  e^{-i q \cdot x}
\  \langle  {\cal N} \vert T\left( J^{\mu \ \dagger} (x)\  J^\nu(0)\right)
\vert  {\cal N}   \rangle \ ,
\label{eq:tprodis}  \eea
where $\vert {\cal N}\rangle $ represents a generic hadronic state and $q$ is the momentum of
the external  vector boson with space-like momentum
 $q^2 \le 0$. In this case, we have
only $q$ as large momentum in the game.
Thus we may write
\bea \label{eq:tmunudis} T^{\mu\nu}&=& -i \int d^4 x \ e^{-i q \cdot x} \
\langle  {\cal N} \vert  \bar \psi(x) \Gamma^\mu S(x-0) \Gamma^\nu \psi(0) 
 \vert  {\cal N}  \rangle  \nonumber \\  &=&  
-i \int d^4 x \  
\langle {\cal N} \vert  \bar \psi(x) \Gamma^\mu S_q(x-0) 
\Gamma^\nu \psi(0) 
 \vert {\cal N} \rangle \ , \eea 
where $\Gamma^\mu=\gamma^\mu$ or $\Gamma^\mu=\gamma^\mu_L$
for electromagnetic or weak charged currents respectively and
$S_q(x-0)= e^{-i q \cdot x} S(x-0)$. By expanding $S_q(x-0)$ in powers of
$-q^2$, as done before for the inclusive decay rate,
we may define  a distribution function  also in this case.
Thus the same formulae apply to the case in which the large momentum is
space-like, as in DIS, or time-like, as in  inclusive  $B$ decays. 
\section{Implementation of the method in numerical simulations}
\label{sec:implementation}
In this section, we briefly explain the method to extract $\tilde f(k_+)$
from the Euclidean lattice correlation functions and discuss the feasibility
of our approach.
Without loss of generality,
we  work with a $B$
meson at rest. The extension of our method to a $B$ meson
with an arbitrary velocity $v$ is
straightforward.\par 
$W^{\mu\nu}(t, \vec Q) $ can be readily obtained from the ratio 
\beq \label{eq:rw} W^{\mu\nu}(t, \vec Q)= \lim_{t_f,t_i \to \infty}
\frac{W^{\mu\nu}_{t_f,t_i}(t, \vec Q) }{S_{t_f,t_i}} e^{- M_B t}\,, \eeq
where
\beq W^{\mu\nu}_{t_f,t_i}(t, \vec Q) = \frac{1}{\pi} \int d^3 x \ e^{- i
\vec Q \cdot \vec x}
\langle 0 \vert   \Phi^\dagger_{\vec p_B=0}(t_f) J^\dagger_\mu(\vec x,
t) 
J_\nu (0)\Phi_{\vec p_B=0}(-t_i)\vert 0 \rangle
, \label{eq:fun}
\eeq
and 
\beq \label{eq:sb}  S_{t_f,t_i}=
\langle 0 \vert  \Phi^\dagger_{\vec p_B=0}(t_f)  \Phi_{\vec p_B=0}(-t_i)
\vert 0 \rangle \, .
\eeq
$\Phi_{\vec p_B}(t)$ is the $B$ interpolating
field with definite spatial momentum $\vec p_B$
\beq
\Phi_{\vec p_B}(t)=\int d^3x e^{-i\vec p_B\cdot\vec x}\Phi_B(\vec x,t)~.
\eeq
\par It is very easy to demonstrate eq.~(\ref{eq:rw}).
In the Euclidean,
using the transfer matrix formalism, we have
\beq 
\Phi_B(\vec x , t)=e^{\hat H t} \Phi_B(\vec x) e^{-\hat H t} \ ,
\eeq
so that the correlation functions have an exponential dependence on the energy
of the external states. 
This implies that, in the limit $t_i,t_f\to\infty$, the
lightest boson state, corresponding to a $\bar B$ meson, dominates the
correlation functions (\ref{eq:fun}) and (\ref{eq:sb}), since all 
higher-energy  
states are exponentially suppressed.
In this limit, with
$t > 0 $, we then obtain
\bea \label{eq:wmunue}
W^{\mu\nu}_{t_f,t_i}(t, \vec Q)\to && \frac{1}{\pi}
\langle 0\vert\Phi^\dagger_B(0)\vert \bar
B(\vec p_B=0)\rangle \times
\left[\int d^3x
\ e^{-i \vec Q \cdot \vec  x} 
\langle B\vert J^\dagger_\mu(\vec x ,t )J_\nu(0)
\vert \bar B \rangle  \right] \times \nonumber \\ &\, &
\langle \bar B(\vec p_B=0)\vert\Phi_B(0)\vert 0\rangle e^{- M_B (t_i+t_f)} \ ,
\eea
and 
\beq \label{eq:sba} 
S_{t_f,t_i} \to
 \langle 0\vert\Phi^\dagger_B(0)\vert \bar B(\vec p_B=0)\rangle
\langle \bar B(\vec p_B=0)\vert\Phi_B(0)\vert 0\rangle 
e^{-  M_B (t_i+t_f)}\, .
\eeq
Thus we have shown that for $t_i,t_f \to \infty$ the ratio in eq.~(\ref{eq:rw})
 directly gives the required quantity and that the knowledge of the
 coupling of the interpolating
field $\Phi_B$ to the physical meson state is not necessary.\par
\par We now discuss the feasibility of the method in actual numerical
simulations. If one uses the complete light quark propagator,
a systematic effect in the extraction of $\tilde f(k_+)$ in 
the relevant region,  $k_+ \sim \bar \Lambda$, is induced
by  contributions of states with  a small invariant mass, 
which dominate the Euclidean correlation functions
 when  $t \to \infty$,  see eq.~(\ref{tproe}).  By injecting
a large momentum $\vec Q$ in the correlation functions, 
the difference between  the  energy $E_X$ of these states 
and the energy $Q_0^+$ of the states we are interested in becomes 
$Q_0^+ - E_X \sim  Q_0^+ - \sqrt{\vec Q^2}
\sim \bar \Lambda$.  Moreover the contribution of these states is suppressed
at least as $\bar \Lambda^2/M_B^2$. In order to compute
$\tilde f(k_+)$, the condition
\beq \label{eq:cod1}{\cal F}=\frac{\bar \Lambda^2}{M^2_B} e^{\bar \Lambda t} \ll 1 \eeq
must then be satisfied.  Thus eq.~(\ref{eq:cod1}) gives a condition on the
maximum time distance which can be used to extract $\tilde f(k_+)$.
We can formulate the condition on the maximum allowed time, in
units of the lattice spacing,  as follows
\beq \label{eq:cod2} N_t = \frac{t}{a} \ll \frac{1}{\bar \Lambda a} \ln(\frac{M_B^2}{\bar
\Lambda^2}) \ . \eeq
By working  with an  arbitrarily small lattice spacing $a$, the 
condition~(\ref{eq:cod2}) can always be satisfied for large values
of $N_t$ and $\tilde f(k_+)$ can then be computed with
negligible uncertainty. In practice, the value of $a$ is limited by
computer resources and we have to worry whether we have a sufficient 
number of points in time  to unfold the shape function.
To give an example, with a lattice spacing $a^{-1}=6$ GeV, a value which
will probably be reached with the 100-gigaflops/teraflops machines already 
available or in construction, a  heavy meson mass $M \sim 2$ GeV and 
$\bar \Lambda=0.3$ GeV, one finds $N_t \sim 75$, which is certainly large
enough.  Note that   $f(k_+)$ cannot be studied for $k_+ \gg \bar
\Lambda$  because, in practice, the correlation function is dominated by subleading
terms in $1/M_B$ at all accessible time distances. 
This region, however, can be studied perturbatively.
\par In order to avoid the limitations on $N_t$, 
a possibility is, of course, to perform the numerical simulations
using the approximate  propagator of
eq.~(\ref{eq:expaSQ}) (we omit the $\Qslash$ in  the numerator because this
gives rise to trivial kinematic factors)
\beq \tilde S_Q = \frac{1}{ Q^2 + 2 i Q \cdot
D + i \epsilon } = \left(
\frac{1}{v\cdot Q} \right) \ \frac{1}{  Q^2/ v \cdot Q +  2 i D_+
+ i \epsilon } \, .  \eeq  
It is straightforward to show that $\tilde S_Q(x)$ can be written as 
\beq \tilde S_Q(x) \equiv \frac{e^{i (v \cdot Q )  x_+ /2} }
{v \cdot Q} S_{\scriptscriptstyle
LEET}(x) \, , \eeq
where $x_+ = n \cdot x$ and $S_{\scriptscriptstyle LEET}(x)$ is
the light-cone   propagator of the Large
Energy Effective Theory (LEET)~\cite{grin}, which satisfies the equation
\beq 2 i D_+ S_{\scriptscriptstyle LEET}(x) =\delta^4(x) \, .\eeq 
Thus, the extraction of  $\tilde f(k_{+})$ from $\tilde S_Q(x)$ is 
equivalent to the use of the LEET.  In our case, we are allowed to use this 
approximation since it has recently  been shown that
the LEET is applicable   to inclusive processes~\cite{ac}, in spite of the 
difficulties  that it may have for exclusive decays~\cite{ugo}.
Note that the calculation of the physical shape function
using the LEET propagator, which is more singular
than the propagator of the full theory, requires a further 
logarithmic renormalisation  of  $W^{\mu\nu}(t,\vec Q)$~\cite{koster},
which can be computed in lattice perturbation theory.
The ultraviolet divergences of the LEET correspond perturbatively
to  infrared divergences in the full theory~\cite{guidoc}. 
In the latter case
the infrared divergences are automatically regularised by the non-perturbative
contributions in the physical matrix elements of the $T$-product of the two
currents and no renormalisation is required.
\par It is not clear to us whether  the use of $S_{\scriptscriptstyle LEET}(x)$ 
will be convenient in practice, since this propagator is much more singular than the
full one. For this reason we expect
that  the  correlation functions computed in numerical simulations
using $S_{\scriptscriptstyle LEET}(x)$ will be affected by larger 
statistical fluctuations.
\par We stress  that in all the formulae derived in this paper, we never
used  the fields of the HQET and, indeed, the same formalism 
 also applies to the calculation of the structure functions in DIS. 
 The reason is that the shape function is 
defined from the OPE in powers of $1/W$ (at lowest order in $1/m_b$)
and one has only to worry about
those terms which become of ${\cal O}(1)$ at the end-point of the
lepton spectrum. Thus, the use of the fields of the HQET would only affect terms of
higher order in $1/m_b$   on which we do not have  control anyway.
In the calculation of the relevant
correlation functions, one may  use the heavy-quark propagator
of  the lattice HQET for the b quark. We recall, however, that
 calculations done using the lattice HQET are afflicted by
considerable difficulties, which are absent in the full 
theory~\cite{wittig,flysac}. Thus, we believe that the best strategy is to
compute the shape function in the full theory, both for the light and 
the heavy quarks, at several values of the
heavy quark mass, and extrapolate the results to the $B$ case. This strategy has been
already successful in the calculation of the heavy meson decay constants,
of the form factors for exclusive semileptonic decays, and of the 
$B^{0}$--$\bar B^{0}$ mixing $B$-parameters~\cite{wittig,flysac}.

\section{Conclusions}
In this paper, we have shown that it is possible to compute,
by numerical simulations of lattice QCD, the shape 
function which describes the lepton spectrum in semileptonic 
$B$ decays and the photon spectrum in radiative $B$ decays. The
same approach can be used for the calculation of the structure 
functions in DIS.  The method is based on first principles and does 
not require  any model assumption. Moreover it avoids 
the calculation of the matrix elements of higher dimensional operators,
which are plagued by power divergences  and the renormalisation of 
which is very difficult to achieve~\cite{sm1}. Indeed, no 
renormalisation is needed and the shape function can be extracted 
directly from suitable ratios of lattice correlation functions. We 
have also proposed  a redefinition of the shape function which avoids 
any reference to $\bar \Lambda$ or to the b-quark mass, and allows us to 
write the differential rate in terms of hadronic quantities only.

\section*{Acknowledgements}
We thank M.~Neubert, C.~Sachrajda and M.~Testa for useful discussions.
We acknowledge partial support by M.U.R.S.T. L.S. acknowledges the support
of German Bundesministerium f\"ur Bildung und Forschung under contract
06 TM 874 and DFG Project Li 519/2-2.

\end{document}